\documentclass[namedreferences]{solarphysics}
\usepackage[optionalrh]{spr-sola-addons} 
\usepackage{epsfig}          
\usepackage{graphicx}        
\usepackage{amssymb}        
\usepackage{color}           
\usepackage{url}             

\newcommand{\sixii}{{Si~{\sc xii}}}
\newcommand{\sixiii}{{Si~{\sc xiii}}}
\newcommand{\sixiv}{{Si~{\sc xiv}}}
\newcommand{\sxv}{{S~{\sc xv}}}

\begin{document}

\begin{article}

\begin{opening}

\title{Silicon Abundance from RESIK Solar Flare Observations}

\author{B.~\surname{Sylwester}$^{1}$\sep
        K.J.H.~\surname{Phillips}$^{2}$\sep
        J.~\surname{Sylwester}$^{1}$\sep
        A.~\surname{K\c{e}pa}$^{1}$
       }
\runningauthor{B. Sylwester {\it et al.}}
\runningtitle{Si Abundance in Flares}

   \institute{$^{1}$ Space Research Centre, Polish Academy of Sciences, 51-622, Kopernika~11, Wroc{\l}aw, Poland
                     email: \url{bs@cbk.pan.wroc.pl}, \url{js@cbk.pan.wroc.pl}, \url{ak@cbk.pan.wroc.pl}\\
              $^{2}$ Mullard Space Science Laboratory, University College London, Holmbury St Mary, Dorking, Surrey RH5 6NT, UK
                     email: \url{kjhp@mssl.ucl.ac.uk} \\
             }

\begin{abstract}
The RESIK instrument on the CORONAS-F spacecraft obtained solar flare and active region X-ray spectra in four channels covering the wavelength range 3.8\,--\,6.1~\AA\ in its operational period between 2001 and 2003. Several highly ionized silicon lines were observed within the range of the long-wavelength channel (5.00\,--\,6.05~\AA). The fluxes of the \sixiv\ Ly-$\beta$ line (5.217~\AA) and the \sixiii\ $1s^2 - 1s3p$ line (5.688~\AA) during 21 flares with optimized pulse-height analyzer settings on RESIK have been analyzed to obtain the silicon abundance relative to hydrogen in flare plasmas. As in previous work, the emitting plasma for each spectrum is assumed to be characterized by a single temperature and emission measure given by the ratio of emission in the two channels of GOES. The silicon abundance is determined to be $A({\rm Si}) = 7.93 \pm .21$ (\sixiv) and $7.89 \pm .13$ (\sixiii) on a logarithmic scale with H = 12. These values, which vary by only very small amounts from flare to flare and times within flares, are $2.6 \pm 1.3$ and $2.4 \pm 0.7$ times the photospheric abundance, and are about a factor of three higher than RESIK measurements during a period of very low activity. There is a suggestion that the Si/S abundance ratio increases from active regions to flares.
\end{abstract}
\end{opening}

\section{Introduction}
     \label{Introduction}

{\it REntgenovsky Spektrometr s Izognutymi Kristalami} (RESIK:  \inlinecite{jsyl05}) was a solar X-ray crystal spectrometer onboard the CORONAS-F spacecraft operating from the time of launch (31 July 2001) until a spacecraft failure terminated the mission in May 2003. The instrument and calibration are described by \inlinecite{jsyl05}. RESIK's wavelength range, 3.40\,--\,6.05~\AA, included spectral lines emitted by highly ionized elements in flare and active region spectra, and also continuum, which unlike many previous crystal spectrometers can be separated from a fluorescence background. A relatively precise absolute calibration makes RESIK ideal for the determination of element abundances with appropriate assumptions for the temperature structure of the emitting plasma. In this article, our motivation has been the determination of the silicon abundance for flares observed by RESIK, and comparison with previous estimates of coronal (both flare and active region) and photospheric silicon abundances.

In previous articles reporting the analysis of RESIK spectra (see \opencite{jsyl12} and references cited therein), we have estimated the abundances of K, Ar, Cl, and S from emission lines of helium-like or hydrogen-like ions of these elements during flares, using a method that involves assigning a ``characteristic temperature" for the emitting plasma equal to that given by the ratio of the two channels of GOES. This  method is applied here for the \sixiv\ Ly-$\beta$, and \sixiii\ $w3$ lines. Although there are other lines in this channel due to \sixiii\ or \sixiv\ transitions, we defer the analysis of their fluxes to later work since the atomic data included in {\sf CHIANTI} are at present under discussion. In Section~\ref{RESIK_instrument} a brief description of the instrument is given, and in Section~\ref{RESIK_observations} the RESIK spectra and their analysis are discussed. In Section~\ref{Abund_Si} the silicon abundance is derived using methods used in our previous work. Comparison with other silicon abundance estimates is made, and the atomic data used for the Si X-ray lines are critically examined, in Section~\ref{Conclusions}.

\section{RESIK Instrument}
\label{RESIK_instrument}

RESIK consisted of two spectrometers, each containing a pair of crystal--detector combinations to give a total of four channels to cover the wavelength range. The crystals were silicon or quartz, having low atomic numbers so that a background, formed by fluorescence of the crystal material by solar X-rays, was relatively small. This is an important advantage over previous spectrometers such as the {\it Bragg Crystal Spectrometer} on {\it Yohkoh}, operational in the 1990s, which had high-$Z$ germanium crystals. The wavelength ranges of RESIK's four channels were  3.40\,--\,3.80~\AA\ (Channel~1), 3.83\,--\,4.27~\AA\ (2), 4.35\,--\,4.86~\AA\ (3), and 5.00\,--\,6.05~\AA\ (4), with a pair of curved Si (111) crystals for Channels 1 and 2 and a pair of quartz ($10\bar10$) crystals for Channels 3 and 4. These ranges included emission lines of hydrogen-like and helium-like Si, S, Cl, Ar, and K and several groups of dielectronic satellite lines. The instrument was uncollimated to maximize its sensitivity (the quoted wavelength ranges are for on-axis sources in the first diffraction order). A careful assessment of instrument parameters \cite{jsyl05} has enabled effective areas to be estimated with a precision of $\approx \pm 20$~\%. Pulse-height analyzers on RESIK allowed the separation, or partial separation, of solar photons from the crystal fluorescence background. Adjustments of pulse-height analyzers over the course of the mission resulted in optimum settings in each channel by 24 December 2002, after which the fluorescence background was entirely removed from Channels 1 and 2 and much reduced for Channels 3 and 4.

\section{RESIK Observations and Analysis}
\label{RESIK_observations}

\begin{figure}
\centerline{\includegraphics[width=0.8\textwidth,clip=,angle=0]{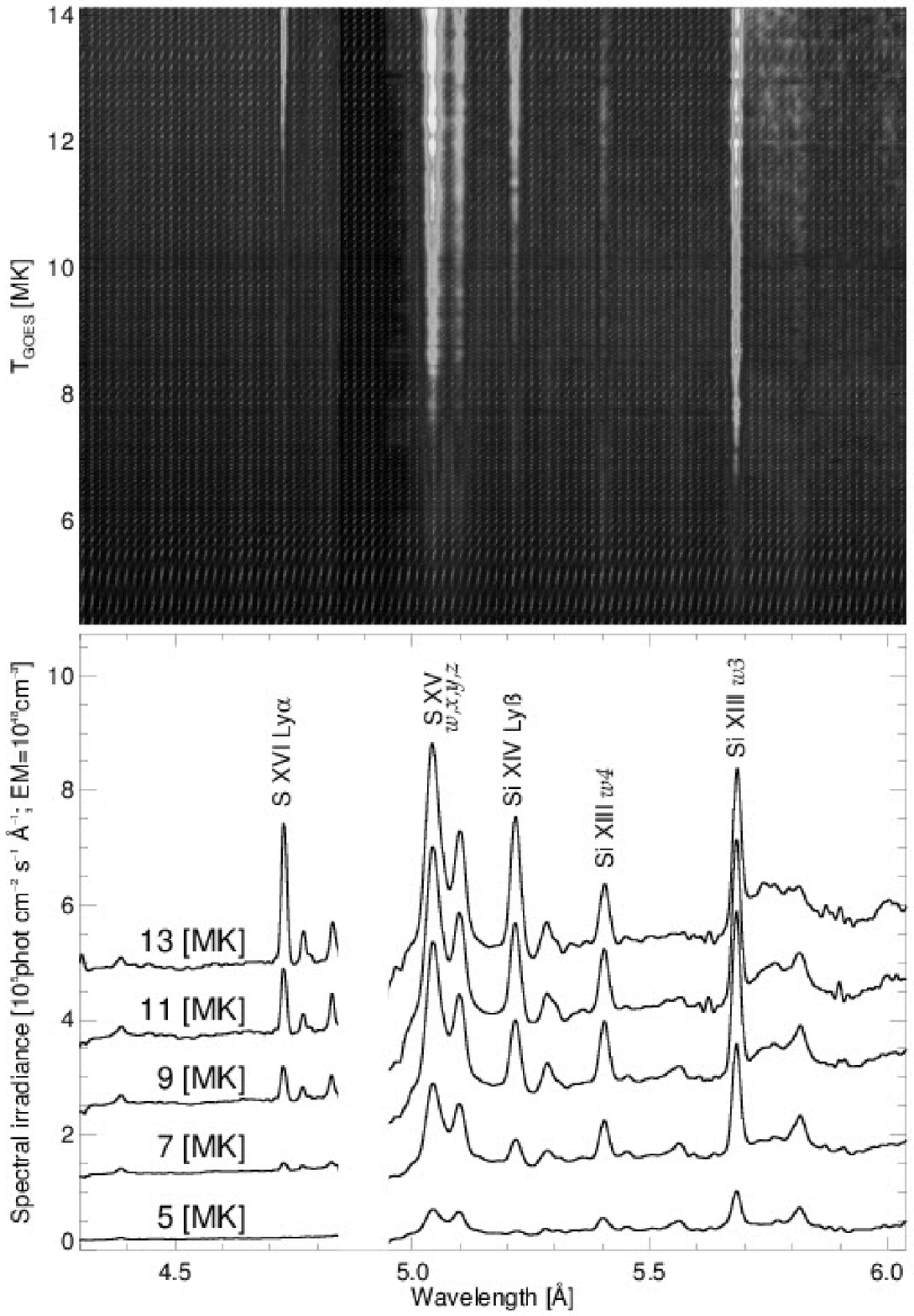}}
\caption{Upper panel: Spectra for Channels 3 and 4 for all 1822 spectra analyzed in this work, stacked in order of $T_{\rm GOES}$ increasing upwards, with the scale shown on the left. Lower panel: Five RESIK Channels~3 and 4 spectra averaged over 2-MK intervals, with the averaged temperature $T_{\rm GOES}$ shown in each case ({\it e.g.} ``5~MK" indicates spectra with $T_{\rm GOES}$ between 4~MK and 6~MK). The \sixiv\ Ly-$\beta$ line at 5.217~\AA\ has increasing intensity relative to the \sixiii\ $1s^2 - 1s3p$ line at 5.688~\AA\ for increasing $T_{\rm GOES}$, while the \sixiii\ line predominates in Channel~4 spectra at lower temperatures. Note that there is a short gap between Channel~3 and Channel~4 spectra (4.86\,--\,5.00~\AA\ for on-axis flares). }\label{RESIK_chans3_4_sp}
\end{figure}

The RESIK spectra analyzed here were taken during the course of 21 flares that, with one exception, were observed after the optimization of the pulse height analyzers on 24 December 2002 so that the fluorescence background in Channels~3 and 4 was reduced as much as possible. The exception is a flare that was observed on 12~November 2002 when the pulse-height analyzers were by chance set to their optimum values for Channels 3 and 4. A total of 1822 spectra were recorded, with data-gathering intervals ranging from two seconds to a few minutes. The data-gathering intervals were inversely related to the incident X-ray fluxes to avoid detector saturation or very low photon count rates with poor statistical significance. Table~\ref{flare_list} gives details of the flares with numbers of spectra collected in each period of observation. For each of these spectra, temperature and emission measure from the emission ratio of the two GOES channels, $T_{\rm GOES}$ and EM$_{\rm GOES}$ (equal to $N_e^2 V$ with $N_e = $ electron density and $V$ the emitting volume), were evaluated. Figure~\ref{RESIK_chans3_4_sp} (upper panel) shows all 1822 Channel~3 and 4 spectra on a colour scale stacked in order of $T_{\rm GOES}$, with the temperature scale shown on the left, while Figure~\ref{RESIK_chans3_4_sp} (lower panel) shows five sample Channel~3 and 4 spectra with the values of $T_{\rm GOES}$ indicated. This figure illustrates how the relative fluxes of the principal lines vary with temperature in accordance with their $G(T)$ or contribution functions, equal to emission per unit emission measure. Thus, in Channel~4 spectra, the \sxv\ lines near 5~\AA\ (temperature of maximum $G(T)$, $T_{\rm max} \approx 16$~MK) and \sixiv\ Ly-$\beta$ line ($T_{\rm max} \approx 16$~MK) are prominent at relatively high temperatures but the \sixiii\ $w3$ line emission ($T_{\rm max} \approx 10.5$~MK) extends to slightly lower temperatures.

\begin{table}
\caption{ Flares Analyzed for Si Abundance. Note: The flare indicated by $^a$ is the first of two overlapping flares.}
\label{flare_list}
\begin{tabular}{rlcccr}
  \hline                   
No. & Date (2002/2003) & Time [UT] of & GOES & Time Range of & Number of  \\
    &     & Flare Peak    & Importance  & Spectra & Spectra \\
  \hline
1 & 12 Nov.  & 07:50 & C5.4 & 07:45\,--\,08:31 & 93 \\
2 & 25 Dec.  & 12:07 & C3.5 & 12:05\,--\,12:33 & 64 \\
3 & 26 Dec.  & 08:35 & C1.9 & 08:26\,--\,08:51 & 53 \\
4 & 7\,-\,8 Jan. & 23:30 & M4.9 & 23:13\,--\,01:07 & 113 \\
5 & 09 Jan.   & 01:39 & C9.8 & 00:14\,--\,02:08 & 297 \\
6 & 21 Jan.  & 02:28 & C8.1 & 02:21\,--\,02:35 &  69 \\
7 & 21 Jan.  & 02:50 & C4.0 & 02:35\,--\,02:44 &  25 \\
8 & 21 Jan.  & 15:26 & M1.9 & 11:57\,--\,18:56 &  290 \\
9 & 25 Jan.  & 18:55 & C4.4 & 18:35\,--\,19:10 & 52 \\
10 & 01 Feb.   & 09:05 & M1.2 & 08:22\,--\,10:14 &  133 \\
11 & 06 Feb.   & 02:11 & C3.4 & 02:01\,--\,02:35 &   34 \\
12 & 14 Feb.  & 02:12 & C5.4 & 01:46\,--\,02:48 &   44 \\
13 & 14 Feb.  & 05:26 & C5.6 & 05:00\,--\,05:57 &   64 \\
14 & 14 Feb.  & 08:47 & C1.4 & 08:42\,--\,09:02 &   23 \\
15 & 14 Feb.  & 09:18 & C5.9$^a$ & 09:11\,--\,09:57 & 42 \\
16 & 15 Feb.  & 08:10 & C4.5 & 06:31\,--\,11:30 &  299 \\
17 & 21 Feb.  & 19:50 & C4.3 & 19:46\,--\,20:10 &   32 \\
18 & 22 Feb.  & 04:50 & B9.6 & 04:40\,--\,05:35 &   19 \\
19 & 22 Feb.  & 09:29 & C5.8 & 08:29\,--\,10:15 &   29 \\
20 & 11 Mar.  & 05:50 & B7.3 & 04:22\,--\,07:23 &   15 \\
21 & 17 Mar.  & 19:05 & X1.5 & 18:53\,--\,19:45 &   32 \\
  \hline
\end{tabular}

\end{table}

\begin{figure}
\centerline{\includegraphics[width=0.8\textwidth,clip=,angle=0]{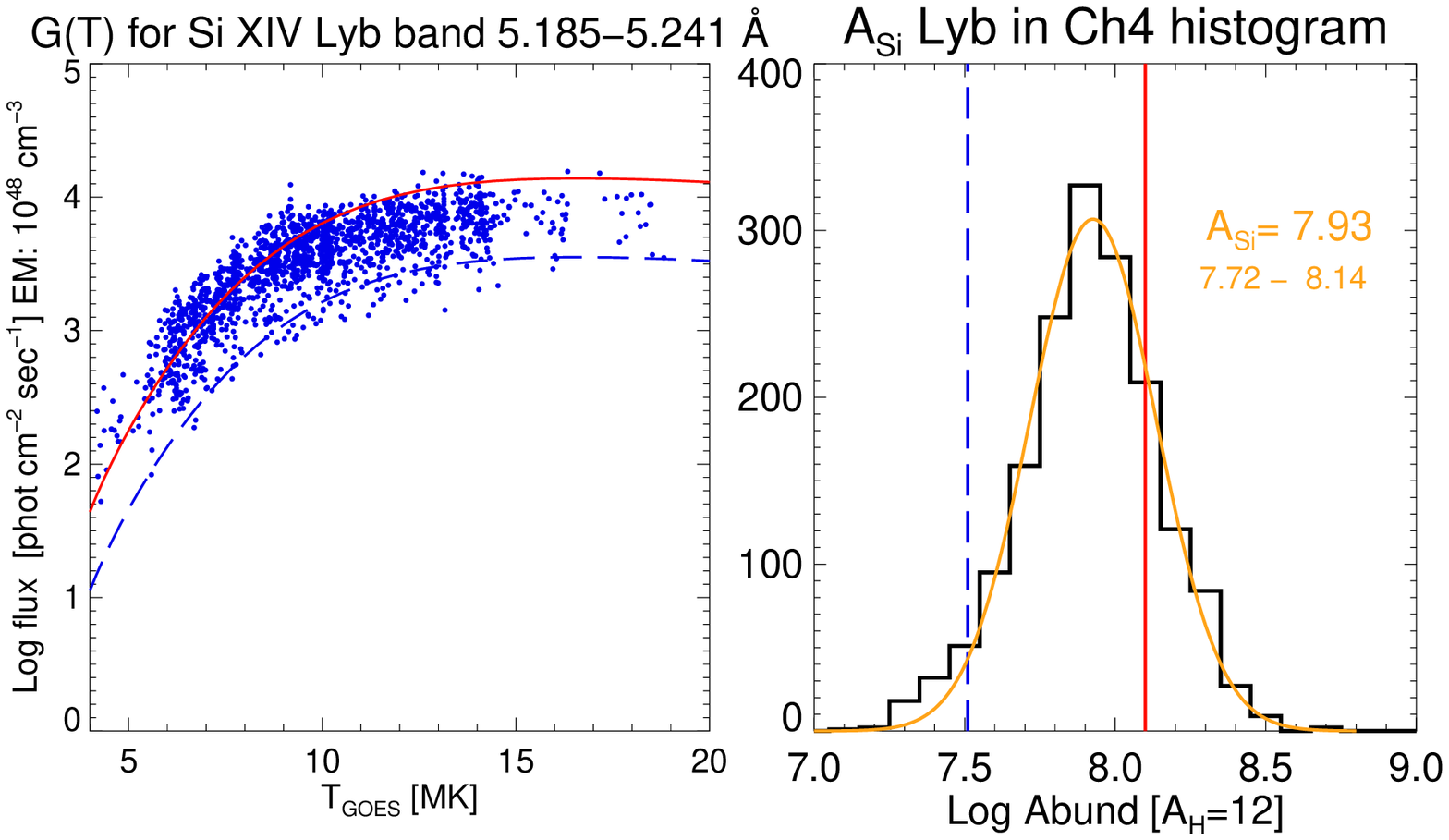}}
\caption{Left panel: Si~{\sc xiv} Ly-$\beta$ line flux [photon cm$^{-2}$ s$^{-1}$, estimated from the photon counts in the range 5.185\,--\,5.241~\AA\ with background in a neighbouring range subtracted] divided by EM$_{\rm GOES}$ for spectra taken during the 21 flares listed in Table~\ref{flare_list} plotted against $T_{\rm GOES}$. 1669 spectra with more than ten photon counts in the Si~{\sc xiv} Ly-$\beta$ line are included. Theoretical $G(T)$ functions are plotted for photospheric (dashed-blue curve) and coronal (continuous-red curve) abundances.  Right panel: Histogram of values of $A({\rm Si})$ derived from the \sixiv\ Ly-$\beta$ line observations plotted in the left panel. The best-fit Gaussian curve indicates a mean abundance of $A({\rm Si}) = 7.92 \pm 0.21$. (The dashed-blue line indicates photospheric Si abundance, the continuous-red line the coronal Si abundance.)}\label{Si_XIV_Lyb}
\end{figure}

RESIK observations of the \sixiv\ Ly-$\beta$ line at 5.217~\AA\ were analyzed by estimating for each spectrum the flux in the wavelength interval 5.185\,--\,5.241~\AA\ and subtracting a neighbouring portion of background, which includes both solar continuum radiation and crystal fluorescence. The values of $T_{\rm GOES}$ and EM$_{\rm GOES}$ at the times of each spectrum were derived from standard software in the {\sf SolarSoft} IDL package, and the line flux divided by EM$_{\rm GOES}$ plotted against $T_{\rm GOES}$; see the left panel of Figure~\ref{Si_XIV_Lyb}. The points at temperatures above $\approx 7$~MK have high statistical quality, but the weakness of the line at lower temperatures prevents measurements at $T_{\rm GOES} \lesssim 5$~MK. The continuous curves in the figure are the theoretical contribution functions $G(T_e)$ from the {\sf CHIANTI} code, defined by

\begin{equation}
G(T_e) =  \frac{N({\rm Si}^{+n}_i)}{N({\rm Si}^{+n})} \frac{N({\rm Si}^{+n})}{N({\rm Si})} \frac{N({\rm Si})}{N({\rm H})} \frac{N({\rm H})}{N_e} \frac{A_{i0}}{N_e} \,\,\,\,\,\,{\rm cm}^3\,\,{\rm s}^{-1}
\end{equation}

\noindent where $N({\rm Si}^{+n}_i)$ is the population of the excited ($i$th) level of the ion Si$^{+n}_i$ ($n=12$ or 13), $N({\rm Si}^{+n})/{N({\rm Si})}$ is the ion fraction as a function of electron temperature [$T_e$], taken in our case from \inlinecite{bry09}. Other quantities in Equation~(1) are $A_{i0}$ (transition probability from level $i$ to the ground state), and $N({\rm H})/N_e$, equal to 0.83. For the abundance of Si relative to H, [$N({\rm Si})/N({\rm H})$], we chose as comparison to the RESIK observed points both the photospheric value given by \inlinecite{asp09}, {\it viz.} $N({\rm Si})/N({\rm H}) = 3.2 \times 10^{-5}$ (equivalent to $A({\rm Si}) = 7.51$) and the coronal value given by \inlinecite{fel00}, $N({\rm Si})/N({\rm H}) = 1.26 \times 10^{-4}$ ($A({\rm Si}) = 8.10$). The red continuous curve in Figure~\ref{Si_XIV_Lyb} is based on the coronal Si abundance, the blue dashed curve on the photospheric Si abundance. Figure~\ref{Si_XIV_Lyb} shows that there is a close correspondence in the shape of the theoretical $G(T_e)$ curves with that defined by the RESIK points, but with a slight tendency for the observed points to be higher than the theoretical curves for $T_{\rm GOES} \lesssim 7$~MK. This discrepancy is discussed further in Section~\ref{Conclusions}.

\begin{figure}
\centerline{\includegraphics[width=0.8\textwidth,clip=,angle=0]{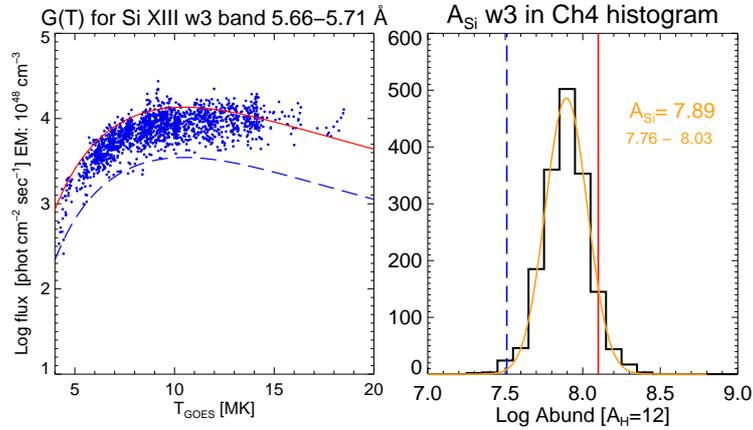}}
\caption{Left panel: \sixiii\ $1s^2 - 1s3p$ ($w3$) line flux [photon cm$^{-2}$ s$^{-1}$, from photon counts in the range 5.66\,--\,5.71~\AA\ with background in a neighbouring range subtracted] divided by $EM_{\rm GOES}$ for the spectra taken during the 21 flares listed in Table~\ref{flare_list}) plotted against $T_{\rm GOES}$. 1684 spectra with more than 50 photon counts in the Si~{\sc xiii} $w3$ line are included. Right panel: Histogram of values of $A({\rm Si})$ from the \sixiii\ $w3$ line observations plotted in the left panel. The best-fit Gaussian curve indicates a mean abundance of $A({\rm Si}) = 7.89 \pm 0.13$. (The dashed-blue line indicates photospheric Si abundance, the continuous-red line the coronal Si abundance.)}\label{Si_XIII_w3}
\end{figure}

The \sixiii\ $w3$ line at 5.688~\AA\ is the most intense line in RESIK's Channel~4 over temperatures ranging from non-flaring active regions \cite{bsyl10} to flares with $T_{\rm GOES}$ up to $\approx 10$~MK. The line flux, taken to be the emission in a band 5.66\,--\,5.71~\AA\ with a portion of the background from a neighbouring band subtracted, is easily measured over this range though at higher temperatures there is extra, unexplained emission between the $w3$ line and a line feature peaking at 5.82~\AA\ due to several \sixii\ dielectronic satellites (transitions $1s^2\,2p\,\,^2P - 1s2p3p\,\,^2D$: \inlinecite{phi06}). Figure~\ref{Si_XIII_w3} (left panel) shows the \sixiii\ $w3$ line emission divided by EM$_{\rm GOES}$ against $T_{\rm GOES}$, with theoretical $G(T_e)$ curves shown, as in Figure~\ref{Si_XIV_Lyb}, for photospheric and coronal Si abundances. The RESIK points follow the same dependence on temperature as the theory curves for $T_{\rm GOES}$ up to $\approx 11$~MK but attain a maximum that is in excess of the maximum of the theory curves. An explanation for this departure is discussed in Section~\ref{Conclusions}.

Shorter-wavelength, weaker lines of \sixiv\ are available for analysis in Channel~3, {\it viz.} the Ly-$\delta$ ($1s - 5p$) and Ly-$\epsilon$ ($1s - 6p$) lines, but we defer this to a later work when the \sixiii\ atomic data become available.

\section{Flare Abundance of Si}
  \label{Abund_Si}

The analysis leading to the abundance of silicon in flare plasmas follows our procedure in previous work, see, {\it e.g.}, \inlinecite{jsyl12} for sulphur lines in RESIK spectra. In brief, for a particular ($i$th) RESIK spectrum, a factor $f_i({\rm Si})$ is estimated given by

\begin{equation}
f_i({\rm Si}) = \frac{F_i}{G(T_i) {\rm EM}_i}
\end{equation}

\noindent where $F_i$ is the flux of the \sixiv\ Ly-$\beta$ or \sixiii\ $w3$ line, the GOES temperature and emission measure at the time of the spectrum are $T_i$ and EM$_i$, and the contribution function [$G$] is evaluated at $T_i$ and with an assumed Si abundance that is either photospheric \cite{asp09} or coronal \cite{fel00}. The measured Si abundance is then given by $f_i$ times the assumed abundance. A histogram is then constructed to show the distribution of the derived abundance values. On a logarithmic abundance scale, the distribution of estimated abundances has been found from our previous work to be close to Gaussian. The peak of the Gaussian distribution gives the most probable abundance estimate and the width the estimated uncertainty.

This has been done for the Si abundances from the \sixiv\ Ly-$\beta$ and \sixiii\ $w3$ lines, as is shown in the right-hand panels of Figures~\ref{Si_XIV_Lyb} and \ref{Si_XIII_w3}. We derive $A({\rm Si}) = 7.93 \pm 0.21$ and $7.89 \pm 0.13$ respectively. The smaller uncertainty of the measurements of \sixiii\ line, which is more intense than the \sixiv\ line for most of the temperature range, reflects the marginally tighter distribution of the observational points in Figure~\ref{Si_XIII_w3}. The slight departure of points at higher temperatures, already noted, does not have an appreciable effect on the Gaussian shape, in particular the peak and width of the curve. This is most likely connected with the accuracy of the atomic data in evaluating $G(T_e)$: the rate coefficients for excitation to the \sixiii\ $n=2$ levels are from the distorted wave calculations of \inlinecite{zha87}, but for higher $n$ the rate coefficients are based on hydrogenic values. There are at present plans to incorporate into the {\sf CHIANTI} data files much more precise rate coefficients from the $R$-matrix calculations of \inlinecite{agg10}. This will enable analysis of the other members of the \sixiii\ lines ($1s^2 - 1s4p$, $1s^2-1s5p$) to be analyzed.There is, then, close agreement of the abundance estimates for these two lines from different ionization stages of Si.

\section{Discussion and Conclusions}
\label{Conclusions}

Figures~\ref{Si_XIV_Lyb} and \ref{Si_XIII_w3} illustrate clearly that the flare abundance of Si from both the \sixiv\ Ly-$\beta$ line and \sixiii\ $w3$ line  is significantly higher than the photospheric abundance. With the latter taken to be $A({\rm Si}) = 7.51 \pm 0.03$ \cite{asp09}, we find the enhancement of the flare value over the photospheric (called the FIP bias by \opencite{fel09}) to be $2.6 \pm 1.3$ (\sixiv) and $2.4 \pm 0.7$ (\sixiii). Our Si abundance estimates are remarkably constant from flare to flare and over the course of the development of each flare as shown by the small scatter of points in Figures~\ref{Si_XIV_Lyb} and \ref{Si_XIII_w3}; Table~\ref{flare_list} shows that for the 21 flares analyzed the GOES peak emission had a range of a factor of $> 400$. This was found to be the case from RESIK measurements for the abundances of at least Ar and S (see \opencite{jsyl12}) and from RHESSI measurements for the abundance of Fe \cite{phiden12}.

Our measurements of flare Si abundances may be compared with coronal abundance estimates from other work, although such estimates are relatively few from  flare spectra. Thus, \inlinecite{vec81} used flare spectra from a crystal spectrometer on OSO-8 to give $A({\rm Si}) = 7.7^{0.2}_{0.3}$, just consistent with our values from the \sixiii\ and \sixiv\ lines. Our values are higher than the two values (7.62, 7.52) measured from \sixiii\ 6.65~\AA\ line emission during flares seen with the {\it Flat Crystal Spectrometer} (FCS) on the {\it Solar Maximum Mission} determined by \inlinecite{flu95}, although it should be noted that the 6.65~\AA\ line falls very near an anomaly in the FCS crystal reflectivity which may affect its intensity.

The analysis of RESIK spectra during a low-activity period in January and February 2003 by \inlinecite{bsyl10} was done by various methods including the derivation of the differential emission measure (DEM). The spectra chosen were from time intervals during the first quarter of 2003, grouped into five GOES level ranges, from A9 to B5. The derived DEM, from the Withbroe--Sylwester method \cite{syl80}, generally had a bimodal distribution, with a low-temperature component (2\,--\,3~MK) and with a high-temperature component (6\,--\,9~MK) which becomes more significant with rising GOES level. An inverse relation of the temperature of the hotter component with GOES level was found, but with $A({\rm Si})$ almost constant, at $\approx 7.4$, about a third of the value obtained here.

Spectra from the {\it Hinode} {\it Extreme-ultraviolet Imaging Spectrometer} (EIS) have been analyzed to give Si/S abundance ratios from quiet-Sun and active region spectra -- the stigmatic nature of the spectrometer allows individual coronal features to be examined. Taking S to be a high-FIP element, {\it i.e.} one with FIP $\gtrsim 10$~eV, the Si/S abundance ratio therefore gives the FIP bias of Si, defined by [Si/S]$_{\rm cor}$/[Si/S]$_{\rm phot}$. The recent analyses by \inlinecite{fel09} and \inlinecite{bro11} have used the flux ratio of Si~{\sc x} and S~{\sc x} lines in EIS spectra to find the Si/S abundance ratio and so the FIP bias of Si. Abundance studies using extreme ultraviolet spectra have a number of difficulties, which are enumerated by \inlinecite{fel09}. Among the most significant is the very crowded nature of this spectral region compared with, say, the RESIK soft X-ray range, so that the possibility of line blends is often high. With this proviso in mind, \inlinecite{fel09} obtain values of Si FIP bias that range from 2.3 (quiet-Sun limb) to 5 (quiet-Sun limb but with ``moderate activity"), and with active region values of 4.6\,--\,4.8. \inlinecite{bro11} observe the active region outflows which have been widely observed with EIS spatially resolved spectra, finding the FIP bias to be 2.5 to 4.1. FIP bias values for the slow solar wind measured by these same authors using data from the SWICS instrument on {\em Advanced Composition Explorer} (ACE) are found to agree with the latter EIS result for a particular active region apparently magnetically connected to the location of {\em ACE}, so indicating the connection of the slow solar wind with the active-region outflow observed by EIS.

The RESIK observations of solar X-ray emission during a period of low activity \cite{bsyl10} include both S and Si lines from which the Si/S abundance ratio can be derived. Although the Si abundance was nearly constant for the five GOES ranges, the S abundance varied from $A({\rm S}) = 6.75$ to 7.25, inversely related to GOES activity level but positively related to the temperature of the hotter component of the bimodal DEM distribution. The Si/S ratio increases from 1.6 to nearly 5 for increasing temperature of the hotter component, similar to the range of Si/S ratios measured by \inlinecite{fel09}. For comparison, the flare Si abundance from the data discussed here and the flare S abundance from \inlinecite{jsyl12} give a Si/S abundance ratio of $5.4^{8.7}_{3.3}$ taking into account the uncertainties of both the Si and S abundance measurements. The Si/S ratios of \inlinecite{bsyl10} may fit a picture in which the Si/S abundance ratio steadily increases in active regions with increasing activity to fully fledged flares as studied here.

In summary, the present analysis of \sixiii\ and \sixiv\ lines seen in RESIK flare spectra give estimates of the Si abundance of $A({\rm Si}) = 7.93 \pm 0.21$ (\sixiv\ Ly-$\beta$ line) and $A({\rm Si}) = 7.89 \pm 0.13$ (\sixiii\ $w3$ line), with very small variation from flare to flare or times within flares, in approximate agreement with measurements by \inlinecite{vec81}. These abundances are a factor $2.6 \pm 1.3$  and $2.4 \pm 0.7$ greater than the photospheric abundance \cite{asp09}. The Si/S abundance ratio for flares as observed by RESIK is large compared with values for non-flaring active regions, suggesting that the ratio depends on the degree of activity.

In future work, we will use improved atomic data, when available in the {\sf CHIANTI} package, to obtain Si abundance estimates from \sixiii\ lines seen in RESIK Channel~3 spectra.

\begin{acks}
Financial support from the Polish Ministry of Education and Science (Grant 2011/01/B/ST9/05861), the European Commissions Seventh Framework Programme (FP7/2007-2013) under the grant agreement eHEROES (project no. 284461), and the UK Royal Society/Polish Academy of Sciences International Joint Project for travel support is gratefully acknowledged. We thank the {\sf CHIANTI} team for their continued support of our work; {\sf CHIANTI}  is a collaborative project involving George Mason University, the University of Michigan (USA), and the University of Cambridge (UK).

\end{acks}


%
%
\bibliographystyle{spr-mp-sola}

\bibliography{RESIK_Si_abund}



\end{article}

\end{document}